\documentclass{article}
\usepackage{spconf,amsmath,graphicx,url,bbm}
\usepackage{amssymb}
\usepackage{tabularx,esdiff,algorithm, mathtools,xparse,subcaption}
\usepackage{algorithmicx}
\usepackage{algpseudocode}

\DeclarePairedDelimiter{\norm}{\lVert}{\rVert}
\usepackage{color}
\usepackage{eso-pic}

\title{Multi-task learning with 
compressible features for Collaborative Intelligence}
\name{{Saeed Ranjbar Alvar and Ivan V. Baji\'c}\thanks{This work was supported in part by NSERC Grant RGPIN-2016-04590.}}
\address{School of Engineering Science, Simon Fraser University, Burnaby, BC, Canada 
}
\begin{document}
%
 \AddToShipoutPicture*{\small \sffamily\raisebox{1.0cm}{\hspace{4.4cm}Copyright 
\copyright \hspace{1mm} 2019 IEEE. The original publication is available for download at ieeexplore.ieee.org.
}}

\maketitle
\begin{abstract}
A promising way to deploy Artificial Intelligence (AI)-based services on mobile devices is to run a part of the AI model (a deep neural network) on the mobile itself, and the rest in the cloud. This is sometimes referred to as \emph{collaborative intelligence}. 
In this framework, intermediate features from the deep network need to be transmitted to the cloud for further processing. We study the case where such features are used for multiple purposes in the cloud (multi-tasking) and where they need to be compressible in order to allow efficient transmission to the cloud. 
To this end, we introduce a new loss function that encourages feature compressibility while improving system performance on multiple tasks. 
Experimental results show that with the compression-friendly loss, one can achieve around 20\% bitrate reduction without sacrificing the performance on several vision-related tasks.  
\end{abstract}
%

\section{Introduction}
\label{sec:intro}
With the advances in Artificial Intelligence (AI), computing, and communications infrastructure, AI-based applications are emerging for mobile devices. The common approach for running the current AI-based mobile applications is to send the data from the mobile device (front-end, or edge) to the cloud (back-end), run AI models in the cloud, and send the results back. New generations of mobile devices are capable of running deep neural models on board~\cite{ai_benchmark}, however, this presents a great challenge due to limited power and computing resources available. 
In~\cite{neurosurgeon}, it is shown that in many cases, the optimal strategy in terms of energy consumption and  computation latency is to split the deep model and distribute the computation between the front-end and the back-end. This paradigm is referred to as \emph{Collaborative Intelligence} (CI).

In CI, the mobile device runs a part of the deep model between the input and some layer,  generates a set of deep features, and sends them to the cloud for further processing by the remainder of the deep model, which resides in the cloud. 
In this context, the issues of deep feature compression~\cite{choi_icip, choi_lossless, battlefiled} and transmission~\cite{DFTS} become important. 
In~\cite{choi_icip}, a HEVC-based lossy deep feature codec was shown to reduce the bitrate by up to 70\% without a drop in performance, compared to uploading images directly to the cloud.  A near-lossless deep feature codec was proposed in~\cite{choi_lossless} and tested on several deep models. In a recent study~\cite{battlefiled}, deep feature coding efficiency of four general-purpose data codecs was examined. In~\cite{DFTS}, a simulator for deep feature transmission over packet-loss channels was presented. All these studies considered only task-specific deep models.

Multi-task (MT) models are expected to offer greater utility for CI because their intermediate features can support multiple tasks in the cloud. However, to our knowledge, MT models were not studied in the context of CI thus far. Some of the recent work on MT models includes~\cite{hyperface}, where a model is developed for simultaneous detection of faces, landmark localization, pose estimation, and gender recognition. Other recent MT models include~\cite{iprivacy}, for detection of privacy-sensitive object classes, and~\cite{multinet}, for joint semantic segmentation, detection, and classification. One issue in training MT models is choosing the loss function that accounts for all the tasks, a problem recently studied in~\cite{cambridge}.     

In this paper we develop a MT learning framework with \emph{compressible features} for CI. 
To accomplish this, we create a loss function that measures compressibility of intermediate features in the model, and then incorporate this term into an overall loss function, so that the model learns not only to perform multiple tasks, but to do so with features that can be efficiently compressed for upload to the cloud. To the best of our knowledge, this is the first work on the topic.  
Section~\ref{sec:proposed} presents the proposed framework and the compression-related loss. Implementation details and experimental results are given in  Section~\ref{sec:Experiments}, followed by conclusions in Section~\ref{sec:conclusion}. 

	

\section{The proposed framework}
\label{sec:proposed}

\subsection{ Multi-task collaborative intelligence}
\label{subsec:multi-task_model}

\begin{figure*}[t!]	
	\centering
	\centerline{\includegraphics[scale=0.30]{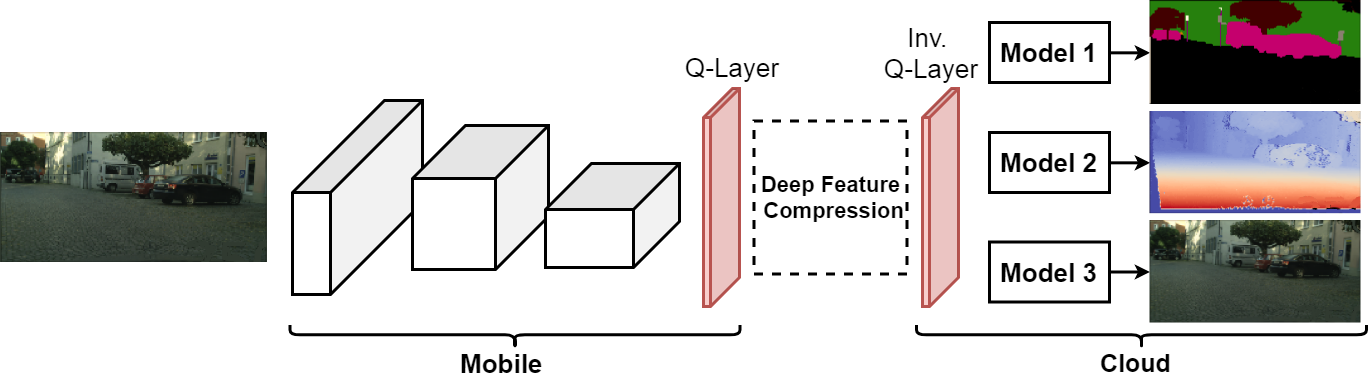}}
	\caption{Multi-task model for collaborative intelligence.} 
	\label{fig:model}
\end{figure*} 

The MT model we consider in this work is shown in Fig.~\ref{fig:model}. It consists of an ``encoder'' part, which consists of layers from ResNet-34~\cite{resnet} and resides on the mobile device. It produces features that will be quantized and compressed, 
and then transferred to the cloud to perform several tasks. 


The inference tasks are executed in the cloud. In Fig.~\ref{fig:model} there are three branches (Models 1-3), each responsible for one of the following tasks: semantic segmentation~\cite{semantic_disparity_ref}, disparity map estimation~\cite{semantic_disparity_ref}, and input reconstruction. All these models consists of convolutional and transpose convolutional (convolution with upsampling) layers, in order to make the resolution of their output equal to the resolution of the input image.  Model 1 maps the received features to a semantic segmentation map, Model 2 maps the features to a disparity map, while Model 3 tries to reconstruct the original input image. We selected these tasks due to the availability of sufficient amount of ground truth, but the overall framework is not dependent on these specific tasks.  

One point to note is that each of the three branches in the cloud uses the same set of features.  
This is in contrast to most of the recent work on MT models~\cite{hyperface,multinet}, where features from different layers of the network are concatenated and then separated according to the tasks to be performed. In the context of CI, such approach would be less appropriate, since features from multiple layers would need to be transferred to the cloud, which would increase the bit rate. In our case (Fig.~\ref{fig:model}), we  only transfer bottleneck features obtained from the last layer of the encoder. This may affect the performance of the our MT model compared to state of the art on each task, but it leads to a more practical CI solution. 

\begin{figure}[b!]	
	\centering
	\centerline{\includegraphics[scale=0.25]{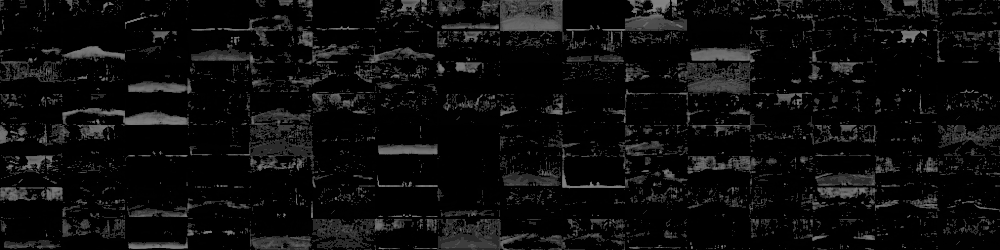}}
	\caption{An example of the tiled quantized deep feature tensor (enhanced for better visualization).} 
	\label{fig:tiled}
\end{figure}

Deep feature compression is accomplished following~\cite{choi_lossless}. The Q-Layer in  Fig.~\ref{fig:model} implements uniform $n$-bit quantization. The quantized feature tensor is then re-arranged into an image tile (Fig.~\ref{fig:tiled}) and compressed using either lossless (e.g. PNG) or lossy (e.g. JPEG) image codec. Note that quantization in the Q-Layer is not differentiable, so during training, similar to the method in~\cite{end-to-end_balle}, uniform noise is added to the Q-Layer input to emulate quantization. This keeps the model differentiable, and it can be trained end-to-end. In the testing phase, quantization is applied as usual. 


\subsection{Deep feature compressibility loss}
\label{subsec:rate_estimation_loss}

Loss functions for the three tasks in Fig.~\ref{fig:model} are well-defined. The loss for semantic segmentation is the cross-entropy loss\ in~\cite{semantic_disparity_ref}. Mean Squared Error (MSE) loss is chosen for depth prediction, and Mean Absolute Error (MAE) loss is chosen for input reconstruction. 
However, to be able to accomplish these tasks with \emph{compressible} features, one needs to include the feature compressibility loss. Such a loss should be related to the bit rate needed to compress the features and it should be differentiable almost everywhere.   


To be more specific, let $\mathbf{F} \in \mathbb{R}^{H \times W \times C}$ be the feature tensor at the output of the last layer at the encoder in Fig.~\ref{fig:model}, where $H$, $W$, and $C$ are the height, width, and channel depth of the feature tensor, respectively. $\mathbf{F}$ is a function of both the input image $\mathbf{X}$ and the weights $\mathbf{W}$ of the encoder model: $\mathbf{F} = f(\mathbf{X}; \mathbf{W})$. Note that $f(\cdot)$, being a forward mapping of a neural network, is an (almost everywhere) differentiable function of the weights $\mathbf{W}$ - otherwise the model would not be able to train via backpropagation. What we want to construct is a loss function $L_r = g(\mathbf{F})$ that is related to the bit rate required for $\mathbf{F}$, and is differentiable with respect to $\mathbf{F}$ almost everywhere. Then, since $L_r = g(f(\mathbf{X}; \mathbf{W}))$, this loss term would also be differentiable with respect to $\mathbf{W}$, which is needed for training.   

One could easily compute the entropy of $\mathbf{F}$ by, say, quantizing its entries, creating a histogram of their values, and then computing the entropy from the normalized histogram. However, this process is not differentiable with respect to the values in $\mathbf{F}$, so we cannot use entropy as the loss term. 
Alternative solutions have been proposed in recent image codecs based on deep networks~\cite{content-weighted-codec,Mentzer,multi-scale-codec}, where the feature value probabilities are predicted using another neural network (hence, in a differentiable manner) such as 
PixelCNN~\cite{PixelCNN} and PixelRNN~\cite{PixelRNN}. Another approach to obtain a bit rate-related quantity is $\rho$-domain analysis~\cite{rho-domain}, where the fraction of non-zero DCT coefficients ($\rho$) is used for this purpose. Authors in~\cite{rate_paper} estimate the rate by calculating the $\ell_1$ norm of the DCT coefficients of either the original (grayscale) input image or the single-channel feature tensor (i.e. feature matrix) obtained from their proposed model. 

In this paper we adopt a similar approach, except that we incorporate spatial prediction into the calculation of the loss. Since spatial prediction is used in high-performance image codecs such as HEVC-Intra, we believe this leads to a better estimate of the rate. 
In addition, our proposed loss can be used on multi-channel tensors as well. 

\setlength{\belowdisplayskip}{2pt} \setlength{\belowdisplayshortskip}{2pt}
\setlength{\abovedisplayskip}{2pt} \setlength{\abovedisplayshortskip}{2pt}
Let $\mathbf{F}_i$ be the $i$-th channel of tensor  $\mathbf{F}$, $i=1,2,...,C$. Spatial prediction is modeled as Differential Pulse Code Modulation (DPCM) along the rows and columns of $\mathbf{F}_i$. Such DPCM can be expressed as matrix multiplication. Let $\mathbf{D}_{H\times H}$ be an $H\times H$ matrix with $1$'s along the main diagonal and $-1$'s along an upper diagonal as shown below:
\begin{equation}
\mathbf{D}_{H\times H}={
\begin{bmatrix}
    1 & -1 & 0 & \dots  & 0 \\
    0 & 1 & -1 & \dots  & 0 \\
    \vdots & \vdots & \vdots & \ddots & \vdots \\
    0 & 0 & 0 & \dots  & -1 \\
    0 & 0 & 0 & \dots  & 1
\end{bmatrix}
}
\end{equation}
Also, let $\mathbf{D}_{W\times W}$ be the $W\times W$ matrix defined analogously. Then horizontal and vertical differencing of $\mathbf{F}_i$ can be expressed as 
\begin{equation}
\label{eq:derivate_x}
    \mathbf{F}^x_i = {\mathbf{F}_i} \mathbf{D}_{W \times W}, \quad 
    \mathbf{F}^y_i = ({{\mathbf{F}_i}^\top} \mathbf{D}_{H \times H})^{\top} = {\mathbf{D}_{H \times H}^\top}{\mathbf{F}_i}
\end{equation}
The horizontal and vertical difference are then averaged, to emulate spatial prediction of elements in $\mathbf{F}_i$ from top and left neighbors: 
\begin{equation}
\label{eq:residual}
    \mathbf{Z}_i = \frac{1}{2} ( \mathbf{F}^y_i +  \mathbf{F}^x_i)
\end{equation}
$\mathbf{Z}_i$ now contains spatial prediction residuals of $\mathbf{F}_i$. Next, DCT is applied to the prediction residual signal as: 
\begin{equation}
\label{eq:transform}
    \mathbf{T}_i= \mathbf{M}_c{\mathbf{Z}_i} \mathbf{M}_r^{\top}
\end{equation}
where $\mathbf{M}_r$ and $\mathbf{M}_c$ are the DCT matrices for row and column transforms, respectively~\cite{dct-matrix}. 
Finally, the feature compressibility loss $L_r$ is computed as:
\begin{align}
\label{eq:L_r}
    L_r = \frac{1}{H \cdot W \cdot C}\hspace{2pt} \sum\limits_{i=1}^{C}\norm{\mathbf{T}_i}_1
\end{align}
This loss is essentially the $\ell_1$ norm of the transformed prediction residual, averaged over the whole feature tensor. Since all the operations used to compute $L_r$ from $\mathbf{F}$ are differentiable almost everywhere, this loss satisfies our requirements and can be used in training. In practice, the derivative of the absolute value at $0$ is set to $1$, to make the derivative well defined everywhere. 

\subsection{Multi-task learning loss}
\label{subsec:multi-task_weights}
The multi-task model in Fig.~\ref{fig:model} is trained end-to-end. For this purpose, a single loss function capturing all task-specific losses is needed. A common way to accomplish this is set the overall multi-task loss as a linear combination of task-specific losses. In our case, we have three actual tasks, and one loss term related to the compressibility of deep features.  

The next question is how to assign weights for these loss terms in the overall loss function. This question has recently been studied in~\cite{cambridge}, where the tasks are divided into segmentation-type tasks and regression-type tasks. Each task-specific loss receives a trainable weight and a correction term that helps suppress trivial solutions where the loss is reduced by setting a task-specific weight to zero. In our case, Task 1 (semantic segmentation) is a segmentation-type task, while other tasks (including compressibility) are treated as regression-type tasks. In this case, the optimal overall loss function (see~\cite{cambridge} for details) is 
\begin{equation}
\begin{split}
\label{eq:total_loss}
L &=  \sum_{i=1}^{3} w_i L_i + w_rL_r + \log{\sqrt{\frac{1}{w_1}}}  \\
& + \sum_{i=2}^{3} \log{\sqrt{\frac{1}{2 w_i}}} + \log{\sqrt{\frac{1}{2 w_r}}} 
\end{split}
\end{equation}
where $L_i$ and $w_i$, $i=1,2,3$, are the task-specific losses and their weights, respectively, while $L_r$ and $w_r$ are the compressibility loss and its weight. Note that the weights $w_i$ and $w_r$ are trainable parameters~\cite{cambridge}, and are updated using the Adam optimizer~\cite{Adam} after each epoch.  

\section{Experimental Results}
\label{sec:Experiments}

\begin{figure*}[tb]	
	\centering
	\centerline{\includegraphics[scale=0.23]{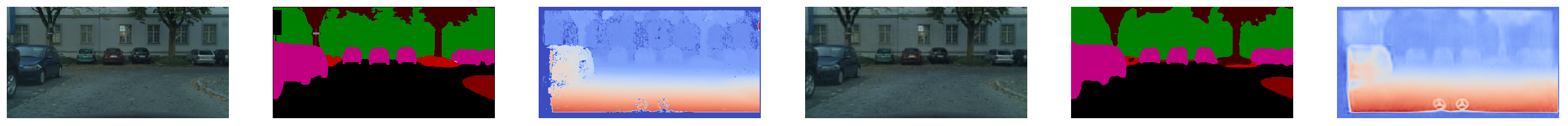}}
	\caption{Left to right: input image, ground-truth semantic segmentation, ground-truth disparity map, reconstructed input, predicted segmentation map, predicted disparity map. 
	} 
	\label{fig:qualitative}
\end{figure*} 

\begin{figure}[t!]
\begin{tabular}{ccc}
\begin{subfigure}{0.45\columnwidth}\centering\includegraphics[width=4.5cm]{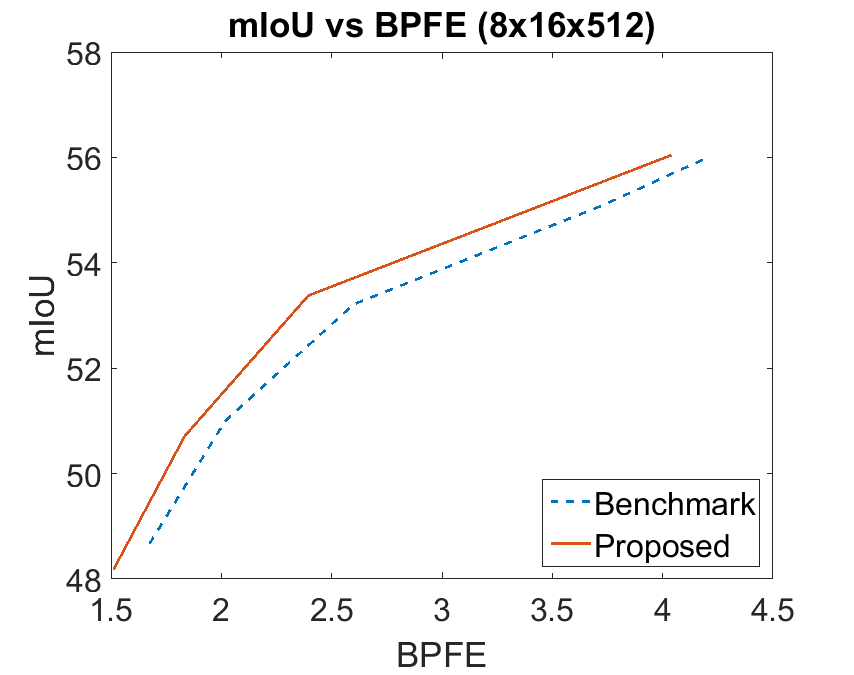}\caption{}\label{fig:curv1}\end{subfigure}&
\begin{subfigure}{0.45\columnwidth}\centering\includegraphics[width=4.5cm]{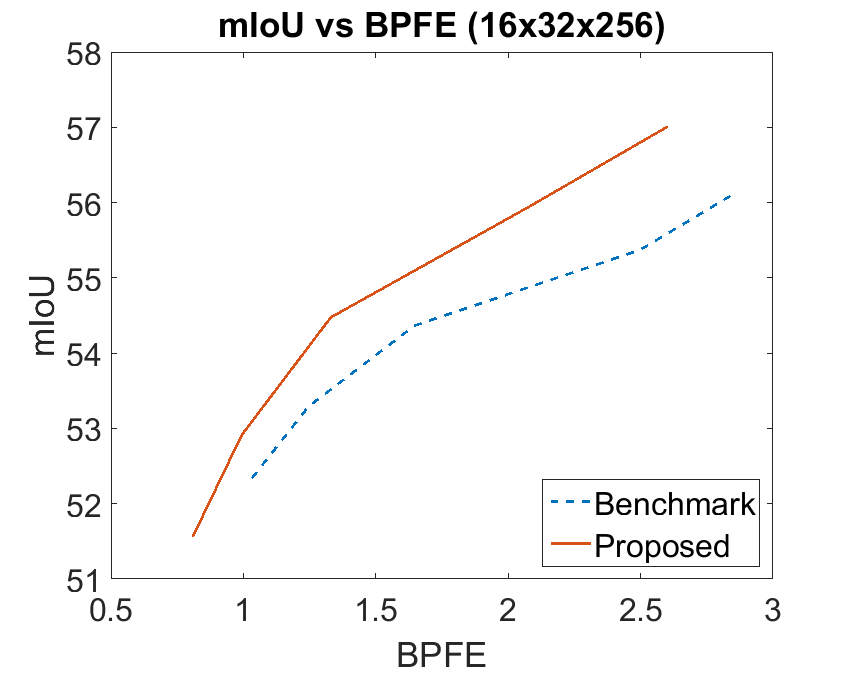}\caption{}\label{fig:curv2}\end{subfigure}\\
\newline
\begin{subfigure}{0.45\columnwidth}\centering\includegraphics[width=4.5cm]{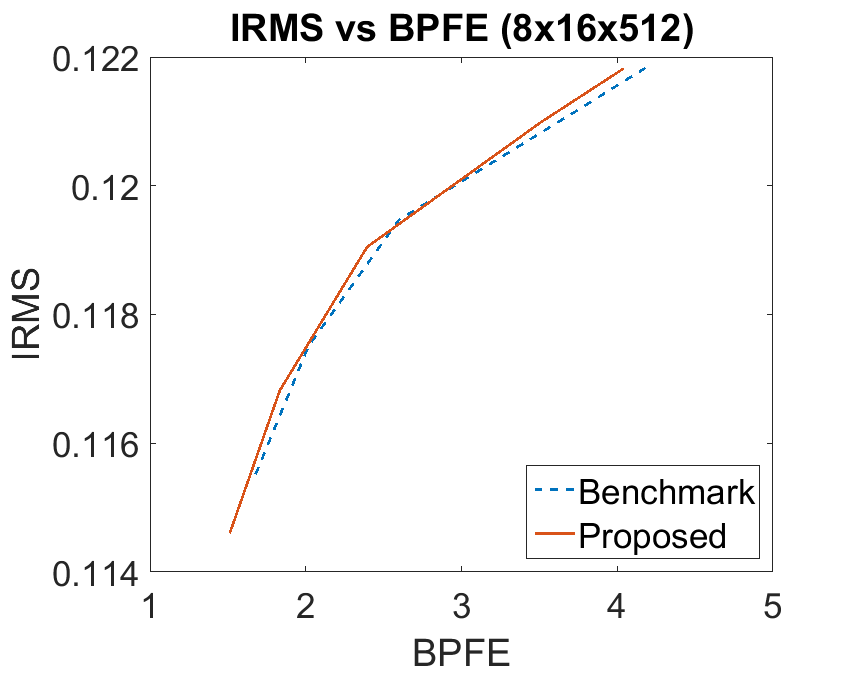}\caption{}\label{fig:curv3}\end{subfigure}&
\begin{subfigure}{0.45\columnwidth}\centering\includegraphics[width=4.5cm]{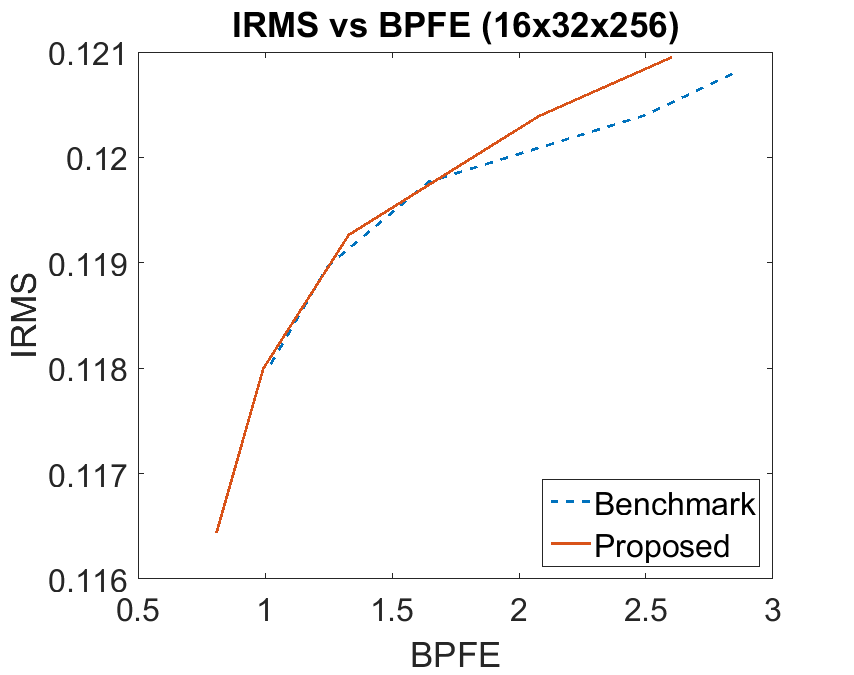}\caption{}\label{fig:curv4}\end{subfigure}\\
\newline
\begin{subfigure}{0.43\columnwidth}\centering\includegraphics[width=4.5cm]{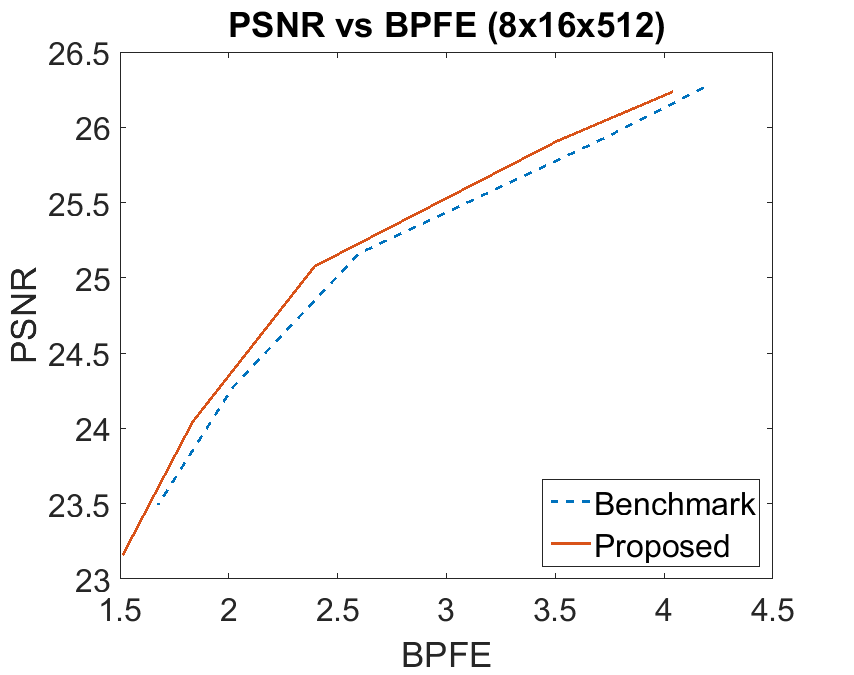}\caption{}\label{fig:curv5}\end{subfigure}&
\begin{subfigure}{0.45\columnwidth}\centering\includegraphics[width=4.5cm]{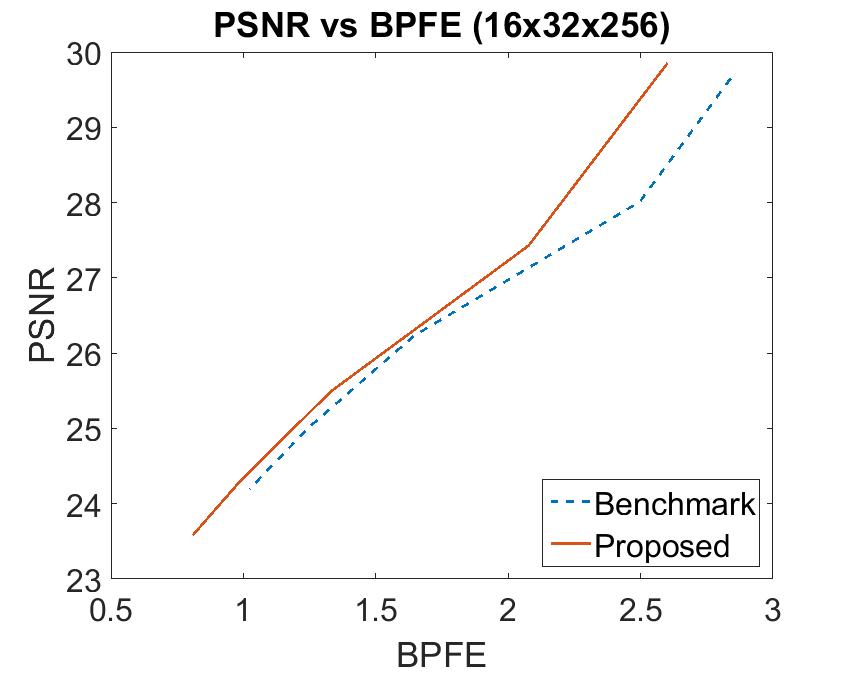}\caption{}\label{fig:curv6}\end{subfigure}\\
\end{tabular}
\caption{The performance curves (task accuracy vs. BPFE) for Encoder 1 in (a), (c), (e), and Encoder2 in (b), (d), (f).}
\label{fig:curve}
\end{figure}

The MT model in Fig.~\ref{fig:model} was implemented in Pytorch~\cite{pytorch}. Cityscapes dataset~\cite{cityscapes} was used for training and testing. Cityscapes has 2,975 training images with the corresponding semantic segmentation and disparity map labels available for each image. Testing was done on the 500 images available in the validation set. To avoid memory limitations on GPU, the resolution of the input images was reduced to $512\times256$. The training was done for 250 epochs with the Adam optimizer~\cite{Adam}, with polynomial decay applied to the initial learning rate of $10^{-3}$. The training batch size is 20. The training and testing was performed on a GeForce GTX 1080 GPU. In the testing phase, the deep features were obtained from the encoder network, quantized, rearranged into a tiled image (Fig.~\ref{fig:tiled}) and then encoded using an image codec, as explained below. 
The reverse process takes place on the cloud side and the decoded features are used for the three tasks. Some visual examples are shown in Fig.~\ref{fig:qualitative}.  

Two encoder networks are used in the experiments. One is the entire ResNet-34 (excluding the top classification layer), which produces  $8\times16\times512$ feature tensors, and which we call Encoder1. The other is ResNet-34 with the last convolutional block excluded (also excluding the top classification layer), which produces feature tensors of size $16\times32\times256$, and which we call Encoder2. For each encoder, a separate inference network is trained for each task.   

The metrics for evaluating the performance of the tasks are mean Intersection over Union (IoU) for segmentation, Inverse Root Mean Square Error (IRMSE) for disparity estimation, and Peak Signal to Noise Ratio (PSNR) for input reconstruction. Since we are interested in the impact of feature compression on these metrics, we show these against Bits Per Feature Element (BPFE) in Fig.~\ref{fig:curve}. Five points are obtained on each curve: one with the lossless PNG codec, and four with lossy JPEG coding with quality parameters 95, 90, 85, 80.  \footnote[1]{Code to reproduce the results in Fig.~\ref{fig:curve} is available at \url{https://github.com/saeedranjbar12/mtlcfci}}  

The benchmark system is the same model in Fig.~\ref{fig:model} but trained without the feature compressibility loss term $L_r$ and its weight $w_r$ in the overall loss function~(\ref{eq:total_loss}). Hence, the comparison with the benchmark shows whether the compressibility loss was useful in improving the system performance. 
As seen in Fig.~\ref{fig:curve}, the proposed system is able to achieve better performance than the benchmark on all tasks at most bit rates. Improvements are especially visible in the mIoU plots for the segmentation task in Fig.~\ref{fig:curve}(a) and (b).


To summarize the differences between performance curves in Fig.~\ref{fig:curve}, we use the Bjontegaard delta (BD) approach~\cite{bjontegaard}, which measures the average bit rate saving for the same performance metric value between two curves. 
The results are shown in Table~\ref{tbl:rate_saving} for both encoders, where the bit rate saving of the proposed method compared to the benchmark is shown. 
Here, negative numbers mean bit rate reduction, which is the desirable outcome. The results show that up to 20\% of bits can be saved on some tasks using an appropriate compressibility loss. 

\begin{table}[t]
{\footnotesize
\begin{tabular}{|c|c|c|c|}
\hline
\begin{tabular}[c]{@{}c@{}}Relative bit-rate \\ reduction \\vs. benchmark\end{tabular} & \begin{tabular}[c]{@{}c@{}}Semantic \\ segmentation\end{tabular} & \begin{tabular}[c]{@{}c@{}}Disparity map\\ estimation\end{tabular} & \begin{tabular}[c]{@{}c@{}}Input\\ reconstruction\end{tabular}  \\ \hline  \hline 
\begin{tabular}[c]{@{}c@{}}Encoder1\\ ($8 \times 16 \times 512$)\end{tabular}        & $-7.81$\%                                                            & $-2.12$\%                                                               & $-4.37$\%           \\ \hline
\begin{tabular}[c]{@{}c@{}}Encoder2\\ ($ 16 \times 32 \times 256 $)\end{tabular}       & $-20.68$\%                                                           & $-5.15$\%                                                               & $-6.09$\%           \\ \hline
\end{tabular}
}
\caption{The average bitrate reduction  over test images of the proposed method compared to the benchmark.}
\label{tbl:rate_saving}
\end{table}

The results in Fig.~\ref{fig:curve} indicate that all three tasks reach a reasonably high performance. However, these numbers are not the current state-of-the-art on each task. As mentioned before, state-of-the-art task-specific models generally avoid bottleneck features and use various connections between multiple layers, making them unsuitable for CI. Our goal here was to study multi-task models in the context of CI, which requires a single set of compressible features to serve all tasks. 

\section{Conclusion}
\label{sec:conclusion}
In this paper we proposed a multi-task learning  framework for collaborative intelligence. We developed a differentiable loss term to measure compressibility of bottleneck features and used it in end-to-end training of a multi-task model. The experimental results show that the proposed approach is able to achieve up to 20\% bit rate savings on some tasks compared to the model trained without the compressibility loss term, without reduction in performance. 
\bibliographystyle{IEEEbib}
\bibliography{refs}

\end{document}